# Fixed field alternating gradient

*Shinji Machida*
ASTeC, STFC Rutherford Appleton Laboratory, Didcot, United Kingdom

**Abstract**
The concept of a fixed field alternating gradient (FFAG) accelerator was invented in the 1950s. Although many studies were carried out up to the late 1960s, there has been relatively little progress until recently, when it received widespread attention as a type of accelerator suitable for very fast acceleration and for generating high-power beams. In this paper, we describe the principles and design procedure of a FFAG accelerator.

## 1  Introduction

The idea of a fixed field alternating gradient (FFAG) accelerator is not new. It was invented in the 1950s right after the alternating gradient (AG) synchrotron came out [1, 2]. Instead of using pulsed magnets as in a synchrotron, FFAG accelerators use constant field magnets like cyclotrons. Unlike cyclotrons, however, FFAGs rely on AG focusing so that the beam size can be much smaller. In the literature [1], there are accelerators described as FFAG betatrons, FFAG cyclotrons and FFAG synchrotrons. This is a little confusing and in these cases, FFAG refers merely to the focusing scheme based on so-called cardinal conditions, which we discuss later. Here we use the terms "FFAG accelerator" or "FFAG" to mean accelerators using the FFAG focusing scheme.

Although there was very active work at the Midwestern Universities Research Association (MURA) from the early 1950s to the late 1960s, the activity stopped when particle physics chose AG synchrotrons for its future tool. The idea of the FFAG was sound, but unfortunately it was not the best accelerator for energy frontier research. An AG synchrotron used more compact magnets and it was easier to obtain high output energy. Nevertheless, the research at MURA associated with the FFAG developments introduced a mathematical formalism and many concepts that became common later in accelerator physics. These include beam stacking, Hamiltonian theory of longitudinal motion, colliding beams, effects of non-linear forces, modelling collective instabilities, use of digital computers in design of orbits, proof of chaotic motion and synchrotron radiation rings. There are two especially informative publications [3, 4] for those interested in a historical view of FFAG developments. The former is free to download.

A new era of FFAG development started in the late 1990s in Japan (there were in fact some activities of FFAG design in 1980s and 1990s in the USA and in Europe; see, for example, Refs. [5, 6]), first in connection with muon acceleration in a neutrino factory [7, 8] and later for wider applications, especially for high-power beam production. The use of fixed field magnets enables very fast acceleration of a beam and is limited only by the available radio-frequency (RF) voltage because there is no restrictive magnet ramping cycle slower than the modulation of the RF frequency. As an extreme example, acceleration of a muon beam, whose lifetime in its rest frame is 2.2 µs, becomes possible although the required RF power would be huge. Such fast acceleration also means that beam acceleration can be repeated more often, for example in a pulsed source. Since beam power is a product of energy, the number of particles per pulse and the repetition rate, it is possible to produce high-power beam via the higher repetition possible in a FFAG accelerator.

A small-scale model of a proton FFAG was first commissioned in 1999 [9] and a prototype of a proton therapy accelerator of 150 MeV energy delivered a beam a few years later. Another FFAG accelerator with similar specifications was constructed as a proton driver test facility linked to the

Japanese Accelerator Driven Subcritical Reactor (ADSR) programme [10]. One of the breakthroughs that made construction of all of these accelerators possible was the success of a novel RF cavity with magnetic alloy material instead of conventional ferrite to modulate the RF frequency. High shunt impedance with very low $Q$ factor is an ideal property for an RF cavity for a FFAG. Activities in Japan revived the potential of the FFAG principle not only as a tool for particle physics but for a variety of applications using state-of-the-art technology.

There was another initiative in FFAG accelerator development outside Japan, which tried to reduce the size of the lattice magnets, specifically for muon acceleration [11]. This line of enquiry led to a new concept being proposed in the late 1990s. Since this FFAG did not follow the so-called scaling law principle of a conventional FFAG, it became named a non-scaling FFAG. An accelerator based on this non-scaling concept was recently built and successfully commissioned in the UK [12].

In the following sections, we briefly explain the principles and design procedure of the different types of FFAG accelerators.

## 2    Scaling FFAG

The guiding field of a weak focusing cyclotron with a single pole decreases gradually with radius. In terms of the magnet field index $n$,

$$n = -\frac{r_0}{B_z(r_0)}\left(\frac{\partial B_z}{\partial r}\right)_{r=r_0} \tag{1}$$

where $r_0$ is the reference radius and $B_z(r_0)$ is the vertical field at the reference radius $r_0$, stable motion in both horizontal and vertical planes requires

$$0 < n < 1 \tag{2}$$

In an AG synchrotron, piecewise magnets of an accelerator lattice have large $n$, but the sign of $n$ alternates, hence the name AG or strong focusing. AG magnets are either combined function type where dipole and quadrupole components co-exist in a single magnet or separated function type where pure dipole magnets are used for bending and quadrupole magnets for focusing. FFAG accelerators use the same the AG principle, but, whereas in an AG synchrotron the magnetic fields are ramped to match the increase in particle momentum under acceleration, in FFAGs the fields are held fixed

Consider the operation of an AG synchrotron without ramping the magnetic fields. It should be possible to increase the beam momentum by applying an RF voltage. The first thing one may observe is the shift of orbit outward or inward depending on the sign of the dispersion function. When the displaced orbit hits one side of the vacuum chamber, the beam is lost. In an ordinary AG synchrotron, this can happen with a momentum increase as small as 1 % or so.

One may also observe a reduction in focusing strength or equivalently transverse tune. The focal length of a quadrupole $f$ is

$$\frac{1}{f} = \frac{B'L}{p/e} \tag{3}$$

where $B'$ is the gradient of the quadrupole magnet, $L$ is its length and $p$ is the beam momentum. When the beam momentum increases, the focusing strength becomes weaker and the focal length becomes longer. This is called a chromaticity effect. Some synchrotrons have chromaticity-correcting sextuples around the ring. The sextupole field profile adds to or subtracts from the field gradient of the lattice quadrupoles depending on the radial position of the beam. When a sextupole is located where the dispersion function is non-zero, off-momentum beams feel either a larger or smaller field gradient and by these means the tune dependence on momentum can be eliminated.

The original FFAG, called a *scaling FFAG*, exhibits a behaviour in terms of orbit and optics similar to a chromaticity-corrected synchrotron without ramping magnets, but designed for much wider momentum acceptance. To avoid the beam hitting an aperture limit, a scaling FFAG simply widens the horizontal size of the vacuum chamber. To increase the momentum range where chromaticity correction works, it uses a magnet whose field gradient depends on radial position as shown in Fig. 1.

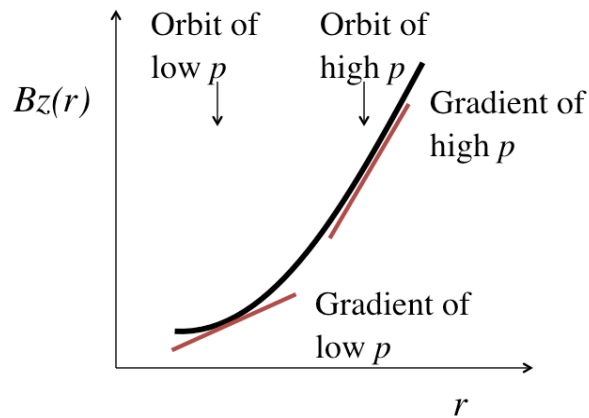

**Fig. 1:** Guiding field profile of a scaling FFAG magnet. The orbit of a high momentum beam sees a stronger field (black curve) as well as a larger field gradient (red curve).

Suppose that a beam with higher momentum circulates around an outer radius and a beam with lower momentum circulates around an inner radius. Since the magnetic field gradient becomes steeper as the beam is accelerated and migrates out, it feels effectively the same focusing force. We define the magnet field index locally as follows where $\rho$ is the bending radius:

$$n = -\frac{\rho}{B}\frac{\partial B_z}{\partial x} \qquad (4)$$

The constancy of *n* independent of beam momentum at each azimuthal position around the ring, expressed as

$$\left.\frac{\partial n}{\partial p}\right|_{\vartheta=const} = 0 \qquad (5)$$

is called one of the *cardinal conditions* of a scaling FFAG.

This is not enough to ensure a constant tune independent of the beam momentum. The second cardinal condition, called geometrical similarity, is written as

$$\left.\frac{\partial}{\partial p}\left(\frac{\rho_0}{\rho}\right)\right|_{\vartheta=const} = 0 \tag{6}$$

This ensures that the momentum-dependent orbits are similar and an exact photographic enlargement of each other.

The derivation above was based on physical intuition, but we can reach the same conclusion starting from the equations of horizontal and vertical betatron oscillations:

$$\frac{d^2x}{ds^2} + \frac{\rho_0^2}{\rho^2}(1-n)x = 0 \tag{7a}$$

$$\frac{d^2z}{ds^2} + \frac{\rho_0^2}{\rho^2}nz = 0 \tag{7b}$$

where $x$ and $z$ are the betatron oscillation amplitudes in the horizontal and vertical directions respectively, and $s$ is the longitudinal coordinate. The requirement to make the coefficients of the restoring force, $\frac{\rho_0^2}{\rho^2}(1-n)$ for horizontal and $\frac{\rho_0^2}{\rho^2}n$ for vertical, independent of the beam momentum leads to the same results as Eqs. (5) and (6).

Now we need to find the magnetic field that satisfies the cardinal conditions. This should have the following form

$$B(r,\theta) = B_0 \left(\frac{r}{r_0}\right)^k F(\vartheta) \tag{8}$$

where $F(\vartheta)$ is a periodic function of a generalized azimuthal angle $\vartheta$. The use of a generalized angle becomes clearer later when a spiral sector type FFAG is discussed. The power $k$ is called the *geometrical magnet field index*. Note that the radial and azimuthal dependence are separated. By expanding the radial dependence in a Taylor series, one can see that the field is a summation of multipoles of all orders:

$$\left(\frac{r}{r_0}\right)^k = 1 + \frac{k}{r_0}(r-r_0) + \frac{k(k-1)}{2!r_0^2}(r-r_0)^2 + \frac{k(k-1)(k-2)}{3!r_0^3}(r-r_0)^3 + \cdots \tag{9}$$

A magnet with this profile gives focusing in one plane, but not in both horizontal and vertical planes simultaneously. To form an AG lattice, we need another magnet with the opposite sign field gradient. In an AG synchrotron, this is done by changing the sign of $k$. In a scaling FFAG accelerator, the sign of $B_0$ is changed as shown in Fig. 2.

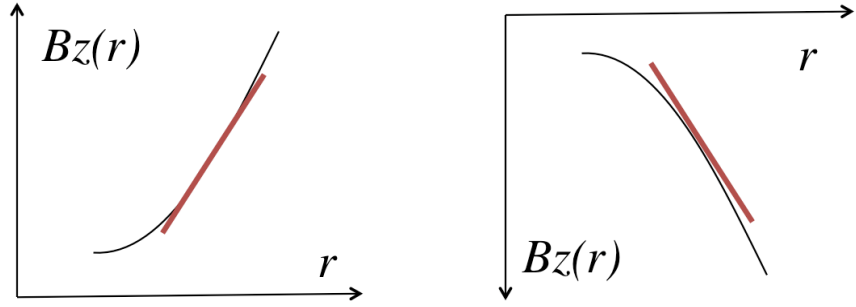

**Fig. 2:** In a FFAG accelerator, the sign of $B_0$ of Eq. (8) changes to realize AG (red lines)

This configuration of a FFAG is called *radial sector type*. The generalized angle is the same as the geometrical angle:

$$F(\vartheta) = F(\theta) \tag{10}$$

You may notice that there is an obvious drawback in doing this. One kind of magnet has the opposite gradient, but at the same time, it has the opposite bending angle as well as shown in Fig. 3. Owing to the cancellation caused by the normal and opposite bending angles, the ring circumference has to be larger than an AG synchrotron which has normal bending only.

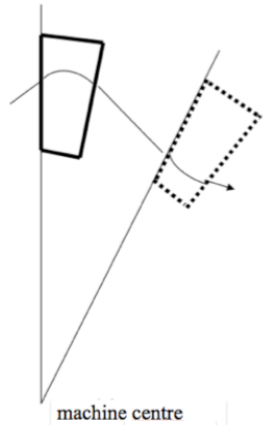

**Fig. 3:** In radial sector type FFAG, magnets with the opposite gradient have opposite bending angles as well

Another way of realizing AG focusing is to introduce edge focusing on one side of a magnet. If a beam does not go into the magnet at right angles to the face, a vertical kick is produced proportional to the vertical displacement:

$$\Delta p_z = -eB_z \tan\zeta \cdot z \tag{11}$$

where $\zeta$ is an injection angle with respect to the normal to the magnet edge. To make the focusing strength independent of momentum, the injection angle of the magnet should satisfy

$$\frac{rd\theta}{dr} = \tan\zeta \tag{12}$$

which then gives a spiral shape. The generalized angle can be defined as

$$\vartheta = \theta - \tan\zeta \cdot \ln\frac{r}{r_0} \qquad (13)$$

In this case the bends can all be in the same direction with no need for alternating signs. This is called a *spiral sector type* as shown in Fig. 4.

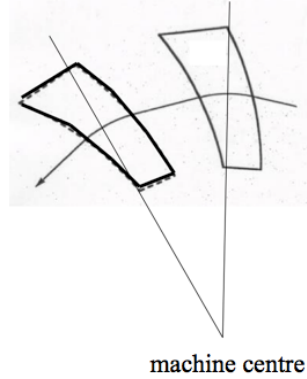

**Fig. 4:** In spiral sector type FFAG, all of the magnets are the same but AG focusing comes from the body and edge

The vertical focusing strength is limited by the need for a practical spiral angle. On the other hand, removing the opposite bending magnets from the lattice makes the circumference smaller than the radial sector type.

You may wonder what is the difference between a FFAG and the Thomas or AVF cyclotron. Apart from the fact that a FFAG does not have isochronous conditions, the Thomas cyclotron is essentially a radial sector FFAG [1]. In that sense, FFAG was not a new invention even when it was proposed in the 1950s. On the other hand, the design strategy is different from that of a cyclotron. A FFAG aims for a much higher momentum accelerator, for example more than 1 GeV, with a reasonable size of magnet by reducing the orbit shift during acceleration. This may not be compatible with the kind of isochronous condition normally found in a cyclotron.

## 3   Linear non-scaling FFAG

Let us consider the operation of an AG synchrotron without ramping magnets again. Remember there were two problems if we did this. One was that a beam would hit the wall due to the dispersion function and the other was that the focusing force would effectively decrease. In a scaling FFAG, we introduced the radial field profile so that the tune was constant for the entire momentum range during acceleration. For the orbit shift, we simply widen the aperture in the horizontal direction.

The orbit excursion in a scaling FFAG is smaller than in a cyclotron, but still not negligible, for example around 0.7 m compared with a 5 m radius for a few hundred megaelectronvolts [10]. This is because of the upper limit of the field index $k$. To squeeze the orbit shift during acceleration, the field index $k$ should be as large as possible. On the other hand, it risks losing the stability inherent in AG focusing because too large a $k$ leads to over-focusing. Also one may notice that most of the orbit shift happens in the lower momentum region where the field gradient is relatively small.

There is another way of designing a FFAG accelerator, which reduces the orbit shift as much as possible, without paying much attention to the tune excursion during acceleration. This is effectively a synchrotron lattice with as small a dispersion function as possible without chromaticity correction. If the dispersion function is small enough, the orbit shift caused by momentum changes can be accommodated in a reasonably sized vacuum chamber. If we eliminate multipoles higher than quadrupole in Eq. (9), the dynamic aperture is expected to be large as well. We choose a quadrupole field gradient that gives a phase advance per focusing unit below 180° at the injection momentum and let it decrease when the beam is accelerated. This is called a *linear non-scaling FFAG*.

A way to design a synchrotron lattice with small dispersion function is well known from the design of synchrotron light sources [13]. The *H*-function is defined as

$$H = X_d^2 + P_d^2 \qquad (14)$$

$$X_d = D/\sqrt{\beta_x} = \sqrt{2J_d}\cos\phi_d \qquad (15)$$

$$P_d = (\alpha_x D + \beta_x D')/\sqrt{\beta_x} = -\sqrt{2J_d}\sin\phi_d \qquad (16)$$

where $D$ and $D'$ are the dispersion function and its derivative, $\beta_x$ and $\alpha_x$ are horizontal Twiss parameters, $J_d$ is invariant in a region without dipole magnets and $\phi_d$ is identical to the betatron phase advance. With bending, the amplitude of $J_d$ changes at a dipole as

$$\Delta X_d = 0 \qquad (17)$$

$$\Delta P_d = \sqrt{\beta_x}\Delta D = \sqrt{\beta_x}\theta \qquad (18)$$

With the help of the *H*-function, we can see that a dipole magnet where the beta function is small makes the dispersion function a minimum. A linear non-scaling FFAG lattice has therefore normal bending at a defocusing quadrupole and opposite bending at a focusing quadrupole [14]. Figure 5 shows the *H*-function and orbits of different momentum of the non-scaling FFAG designed in Ref. [14].

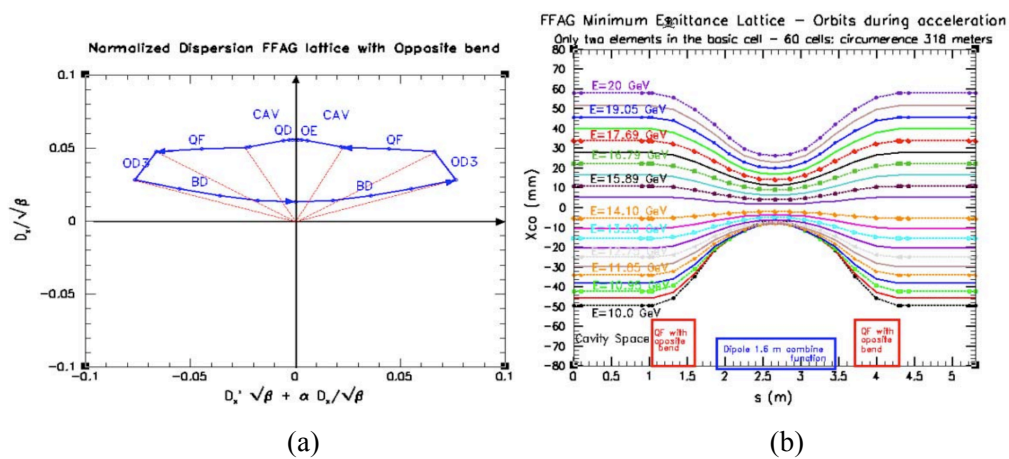

**Fig. 5:** (a) Normalized dispersion space: the opposite bend at QF quadrupole, while the major bend is at BD. (b) Closed orbit offsets during acceleration from 10 GeV to 20 GeV. (Reproduced from Ref. [14].)

Although the design avoids at least a systematic resonance at 180° phase advance per unit cell, a potential problem is crossing higher-order resonances and non-systematic resonances during acceleration as shown in Fig. 6. In the early days, it was thought that keeping transverse tunes independent of momentum during acceleration was the essential design requirement of an accelerator with the AG principle. Resonances excited by repetitive actions to the beam eventually lead to beam loss. A linear non-scaling FFAG challenges this conventional wisdom.

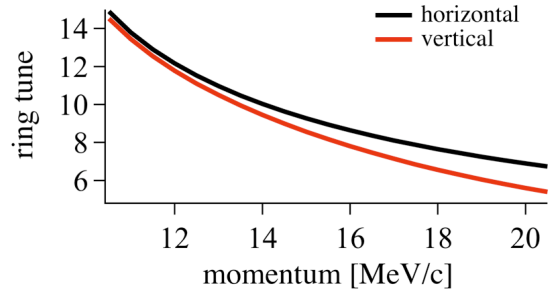

**Fig. 6:** Tune excursion in a linear non-scaling FFAG. Although the phase advance per unit cell is chosen below 180°, the total tune of the ring (in this example 42 unit cells) should cross several integer tunes during acceleration.

The effect of resonance crossing depends on the speed of crossing and the strength of the resonances. For some applications such as muon acceleration, the acceleration is very fast so that the fast tune excursion probably makes the resonance crossing harmless. In fact muon acceleration is the original application for which the linear non-scaling FFAG was designed. On the other hand, when we use a linear non-scaling FFAG for ordinary applications such as a particle therapy accelerator, the beam circulates for of the order of 1000 to 10 000 turns and therefore the speed of resonance crossing would be slow. We do not know how slow a resonance crossing could be tolerated in a linear non-scaling FFAG. This is an on-going research subject.

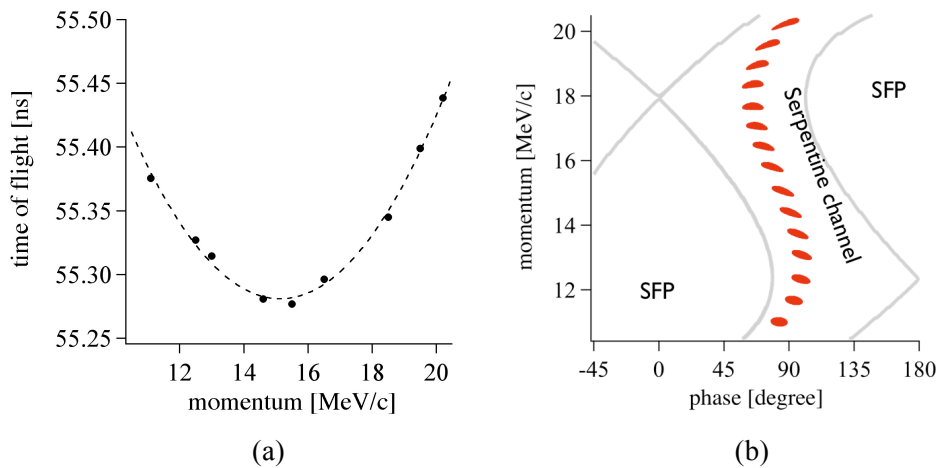

**Fig. 7:** (a) Orbital period (or time of flight) in a linear non-scaling FFAG as a function of momentum. (b) Serpentine channel created between two RF buckets with stable fixed points (SFP) shown. Red marks indicate a beam injected at the lower momentum and accelerated to the top.

A small orbit shift for the entire momentum range also means the orbital length does not vary much as a function of momentum. If a beam is in the ultra-relativistic regime, it also means that the orbital period is almost constant. This enables the use of a constant RF frequency cavity if the phase slip during acceleration is small enough. In fact, in muon acceleration, which is completed in about 10 turns, a beam stays near the crest of the RF voltage throughout. By adjusting the lattice so that the orbital period during the acceleration has a parabolic shape as shown in Fig. 7(a) and choosing a RF frequency that corresponds to the orbital period within the parabola, we can create RF buckets close to each other with slightly different momentum centres. Instead of injecting a beam inside the RF bucket, the channel between the two RF buckets can be used for acceleration. This is called *serpentine channel acceleration*. This novel acceleration scheme with stability in a linear non-scaling FFAG was demonstrated recently in the UK [12].

## 4 Other types of FFAG

The non-scaling FFAG is a promising idea, but it is not clear whether the resonance crossing during acceleration is harmless. There are more concerns when we design an accelerator with moderate acceleration rate with moderate RF voltage such as one for particle therapy. In fact, a simulation study showed that, at the very least, more accurate alignment is required when we accelerate a beam at a slower rate in a linear non-scaling FFAG [15].

It is possible to fix a tune without following the cardinal conditions. One way is to start with the design of a scaling FFAG and relax the conditions. For example, we can truncate Eq. (9) at some particular order. Lattice magnets can be rectangular in shape instead of radial sector type. In these magnets, equipotential lines are straight rather than arcs of a circle. Within a unit cell, three magnets forming triplet focusing are aligned along a line instead of a circle. With careful design of the field profile in each magnet, it is still possible to ensure the tune variation is negligible over a wide range of momentum [16]. This design principle does not follow the scaling rule, but it does not use linear quadrupole magnets either. Therefore, it is called a *non-linear non-scaling* FFAG.

We can fix the tune, based on a linear non-scaling FFAG design, by introducing non-linear magnetic fields. It is like introducing sextupole magnets to correct chromaticity in a synchrotron lattice. Sextupole components are not enough, however, to correct the tune over a wide momentum range. Higher-order non-linearities such as octupole and/or decapole are required. This is another approach to designing a non-linear non-scaling FFAG.

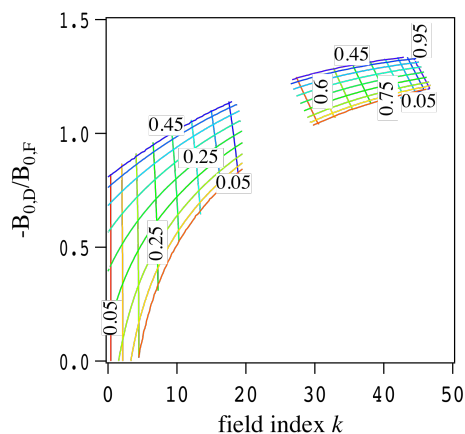

**Fig. 8:** Stability diagram with a practical lattice configuration with the geometrical field index $k$ as abscissa and the ratio of focusing and defocusing magnet strengths as ordinate. Upright numbers indicate vertical cell tune and vertically aligned numbers indicate horizontal cell tune. Lines are drawn with 0.05 step. (Reproduced from Ref. [17].)

One drawback of a scaling FFAG compared with a non-scaling FFAG is its relatively large orbit shift during acceleration. Addressing this was the primary concern that led to the invention of the non-scaling FFAG. There is another way to reduce the orbit shift, however, while keeping the scaling principle. It was well known that there is another stability region in phase space when we increase the focusing strength and produce phase advances over 180° per cell, as shown in Fig. 8. In a synchrotron lattice, there is no advantage in choosing the second stability region because the beta function becomes larger and alignment tolerance and manufacturing specifications become tighter. On the other hand, the larger gradient needed to produce stronger focusing reduces the orbit shift in a FFAG. If we could maintain a reasonable tolerance in the design using the second stability region, it would be a big advantage. With a triplet focusing structure, it turns out that we can reduce the orbit shift by a factor of five if we use the second stability region [17]. The dynamic aperture is smaller and tolerance is tighter, but they are at manageable practical levels.

So far, we have discussed the type of optics for a FFAG used as a circulating accelerator. The optics that can accommodate large momentum spread can be used for beam transport as well. It has been found that such a beam transport line can be realized as the limit of a ring with infinitely large bending angle [18, 19] as shown in Fig. 9(a). A dispersion suppressor at the end of the transport line may add another advantage. Owing to its wide momentum acceptance, this system can be applied to the gantry of a particle therapy facility or a dump line for a synchrotron where the tripped beam momentum could be any momentum from injection to extraction.

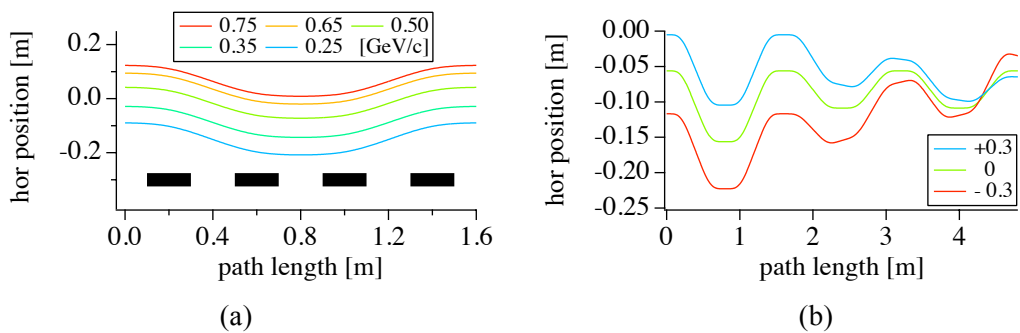

**Fig. 9:** (a) Different momentum orbits in a unit cell which satisfies the periodic boundary condition. Rectangles at the bottom show the position of FDDF magnets. (b) Different momentum orbits in a normal cell with dispersion suppressor. Numbers in the legend show the momentum deviation from the reference momentum. The normal cell is from 0 m to 1.6 m and the dispersion suppressor section from 1.6 m to 4.8 m. (Reproduced from Ref. [18].)

## Acknowledgments

The author would like to thank C. R. Prior and D. J. Kelliher for their careful reading and comments on this lecture note.

## References

[1] K.R. Symon, D.W. Kerst, L.W. Jones, L.J. Laslett and K.M. Terwilliger, *Phys. Rev.* **103** (1956) 1837.
[2] A.A. Kolomensky and A.N. Lebedev, *Theory of Cyclic Accelerators* (North-Holland, Amsterdam, 1966).
[3] F.T. Cole, O Camelot! Memoirs of the MURA years, http://accelconf.web.cern.ch/accelconf/c01/cyc2001/extra/Cole.pdf (1994).

[4] L. Jones, F. Mills, A. Sessler, K. Symon, D. Young, *Innovation Was Not Enough* (World Scientific, 2009).

[5] T.K. Khoe and R.L. Kustom, *IEEE Trans. Nucl. Sci.* **30** (1983) 2086.

[6] R. Kustom, S.A. Martin, P.F. Meads, G. Wuestefeld, E. Zaplatin and K. Ziegler, Proc. of the European Particle Accelerator Conference, 1994, p. 574.

[7] F. Mills and C. Johnstone, Proc. 4th International Conference Physics Potential and Development of $\mu^+\mu^-$ Colliders, Transparency Book (1997).

[8] S. Machida, *Nucl. Instrum. Methods Phys. Res., Sect. A* **503** (2003) 41.

[9] M. Aiba, *et al.*, Proc. of the European Particle Accelerator Conference, 2000, p. 581.

[10] M. Tanigaki, *et al.*, Proc. of the European Particle Accelerator Conference, 2006, p. 2367.

[11] C. Johnstone, W. Wan and A. Garren, Proc. of the Particle Accelerator Conference, 1999, p. 3068.

[12] S. Machida, *et al.*, *Nature Phys.* **8** (2012) 243.

[13] D. Trbojevic and E. Courant, Proc. of the European Particle Accelerator Conference, 1994, p. 1000.

[14] D. Trbojevic, J.S. Berg, M. Blaskiewicz, E.D. Courant, R. Palmer and A. Garren, Proc. of the Particle Accelerator Conference, 2003, p. 1816.

[15] S. Machida, *Phys. Rev. ST Accel. Beams* **11** 094003 (2008).

[16] S.L. Sheehy, K.J. Peach, H. Witte, D.J. Kelliher, S. Machida, *Phys. Rev. ST Accel. Beams* **13** 040101 (2010).

[17] S. Machida, *Phys. Rev. Lett.* **103** 164801 (2009).

[18] S. Machida and R. Fenning, *Phys. Rev. ST Accel. Beams* **13** 084001 (2010).

[19] Y. Mori, T. Planche and J.B. Lagrange, Proc. of FFAG Workshop '08J, http://hadron.kek.jp/FFAG/FFAG08J_HP/ (2008).